# TECHNOCRATIC MODEL OF THE HUMAN AUDITORY SYSTEM

O. PALAGIN, M. SEMOTIUK

**Abstract**

In this work, we investigate the phenomenon of transverse resonance and transverse standing waves that occur within the cochlea of living organisms. It is demonstrated that the predisposing factor for their occurrence is the cochlear shape, which resembles a conical acoustic tube coiled into a spiral and exhibits non-uniformities on its internal surface. This cochlear structure facilitates the analysis of constituent sound signals akin to a spectrum analyzer, with a corresponding interpretation of the physical processes occurring in the auditory system. Additionally, we conclude that the cochlear duct's scala media, composed of a system of membranes and the organ of Corti, functions primarily as an information collection and amplification system along the cochlear spiral. Collectively, these findings enable the development of a novel, highly realistic wave model of the auditory system in living organisms based on a technocratic approach within the scientific context.

Keywords: cochlea, Reissner's membrane, basilar membrane, transverse resonance, elastic wave, traveling wave, standing wave, shear wave, frequency, frequency difference, pressure, inner ear, middle ear.

**Table of Contents**

Alexandr Palagin
Department of Microprocessor Engineering
Institute of Cybernetics of NASU
Kyiv, Ukraine    palagin_a@ukr.net

Miroslav Semotiuk
Department of Microprocessor Engineering
Institute of Cybernetics of NASU
Kyiv, Ukraine       semo@i.ua

## Introduction

The relationships between the objective characteristics of sound and the sensations of its perception are usually established based on subjective observations and are the subject of physiological research. These relationships are only occasionally described using empirical formulas, more often represented in the form of graphs, and sometimes they are purely descriptive. The empirical formulas themselves are obtained using a black-box approach, so they do not necessarily reflect the essence of the processes that occur in the ear. Despite the mechanical simplicity of the cochlear structure, there is still no consensus on how the acoustic signal is transmitted from the stapes to the auditory analyzer and how it is analyzed. Most theories of hearing focus on the waves that arise, or whether they arise at all, and remain contentious. In other words, there is currently no single reliable theory that explains all aspects of human sound perception. This is because the authors of various theories attempt to explain the function of only specific parts of the ear, such as the organ of Corti or the middle ear, independently of other components. Moreover, these theories often do not consider the coordination of input and output parameters of these components or their shapes. However, the ear is ultimately a physical acoustic device, even though it is constructed from living tissues, and it should follow the laws of acoustics.

The question arises: how does this device appear to the eyes of an acoustic engineer? The authors of this work attempt to answer this question by proposing a technocratic model of human hearing. In doing so, they do not consider the physiological aspects of hearing that are commonly accepted in most theories. Instead, they focus solely on the physics of the processes that occur in the ear.

## 1. Structure of the Human Ear

To present the next material, we need to discuss the structure of the human ear. Let's use information from open sources without claiming novelty. The ear is an organ in vertebrate animals, including humans, that functions for the perception of sound and maintaining spatial balance. The human ear is capable of perceiving sound waves with lengths ranging from 1.6 cm to 20 m, corresponding to frequencies of 16-20,000 Hz (vibrations per second). The human ear and those of other mammals consist of three parts: the outer, middle, and inner ear. The outer and middle ears are relatively simple in structure, and their function is to transmit and amplify sound vibrations. The inner ear is the most complex in structure and, in addition to sound perception, also provides balance and a sense of body position in space [1].

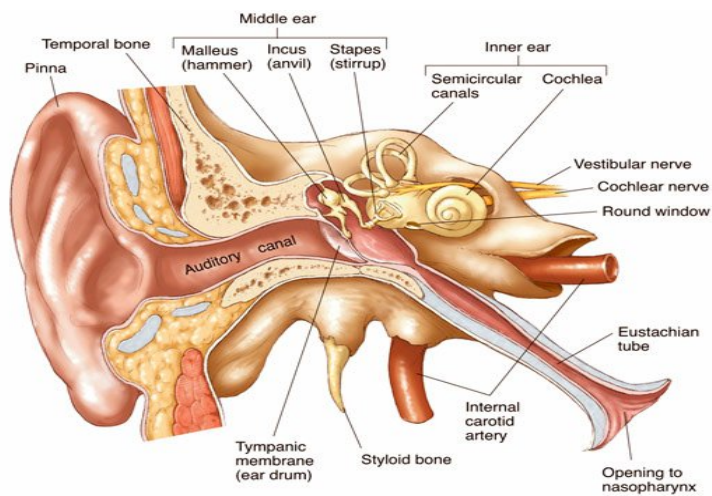

Figure 1 - Structure of the Ear

*The external ear* consists of the auricle (also known as the pinna) and the external auditory canal. The auricle surrounds the external auditory canal and is composed of elastic cartilage covered by skin. Some parts of the auricle, such as the earlobe, lack cartilaginous tissue. The function of the auricle is to direct sound into the external auditory canal [2].

*The external auditory canal* is a short, curved tube (2.5 cm in length and 0.6 cm in diameter) that leads to the eardrum. Near the auricle, the external auditory canal is supported by elastic cartilage, while the rest of it passes through the temporal bone. The entire canal is lined with skin and contains short hairs, which are necessary to trap various foreign objects, such as dust and small insects [2].

*The tympanic membrane,* located at the boundary between the external and middle ear, is a thin, transparent membrane made of fibrous connective tissue. It is covered with skin on its external side and mucous membrane on its internal side. The fibers of connective tissue are arranged predominantly in a circular pattern around the periphery of the membrane and radially in the center. The tympanic membrane has the shape of a flattened cone, with its apex directed toward the cavity of the middle ear. It is attached by a fibrous-cartilaginous ring to the tympanic sulcus. The tympanic membrane vibrates in response to sound waves and transmits these vibrations to the auditory ossicles of the middle ear [2].

*The middle ear,* also known as the tympanic cavity, is a small, air-filled space located within the temporal bone of the skull. It is lined with mucous membrane and is bounded by the tympanic membrane on one side and the bony wall with two openings (oval and round windows leading to the inner ear) on the other side. The upper portion of the middle ear forms a bony, arch-like structure known as the epitympanic recess. On the medial wall of the middle ear is a nipple-like protrusion known as the promontory, which allows it to make contact with the mastoid cells of the temporal bone [5].

*The lower part* of the tympanic cavity contains the opening of the auditory (pharyngotympanic or Eustachian) tube, which extends downward and connects the middle ear cavity to the nasopharynx (the part of the throat that is behind the nasal passages). Most of the time, this auditory tube is flattened and closed. It only opens during actions like yawning or swallowing to equalize the pressure inside the middle ear with the external atmospheric pressure. This equalization is essential to allow the tympanic membrane to vibrate normally. If there's a pressure difference between the outside and inside of the tympanic cavity, it can lead to the sensation of "plugged ears," often experienced during rapid changes in altitude, such as during takeoff and landing in an aircraft. The middle ear cavity contains three auditory ossicles: the malleus (hammer), incus (anvil), and stapes (stirrup). The handle of the malleus is attached to the tympanic membrane, while the base of the stapes is connected to the oval window of the inner ear. These auditory ossicles are supported by ligaments that originate from the walls of the tympanic cavity and are connected to each other by synovial joints. The primary role of the auditory ossicles is to transmit the vibrations from the tympanic membrane to the oval window. Since the tympanic membrane has approximately 22 times the surface area of the oval window, there is significant amplification of the vibrations. Additionally, within the tympanic cavity, there are two small skeletal muscles: the tensor tympani muscle, which originates from the wall of the auditory tube and attaches to the malleus, and the stapedius muscle, which is stretched between the posterior wall of the tympanic cavity and the stapes. When the ear is exposed to very loud sounds, these muscles reflexively contract, reducing the vibrations of the oval window of the inner ear. This reflex helps protect the delicate hair cells of the cochlea from damage due to loud noises [2].

*The inner ear,* due to its complex shape, is also referred to as the labyrinth. It is situated deep within the temporal bone, located behind the eye sockets. The inner ear consists of two main parts. *The bony labyrinth* comprises a system of winding canals within the temporal bone and is filled with perilymph, a fluid similar in composition to cerebrospinal fluid. These canals can interconnect through the helicotrema.

*The membranous labyrinth* is a series of membranous sacs and ducts contained within the bony labyrinth. It is filled with endolymph, which is chemically similar to intracellular fluid and rich in K+ ions. The formation of endolymph involves the endolymphatic sac, connected to the rest of the membranous labyrinth by the endo-lymphatic duct [2,..9]. The inner ear comprises three main sections: the vestibule (utricle and saccule), the semicircular canals, and the cochlea.

*The vestibule,* also known as the vestibular system or vestibular apparatus, is the central egg-shaped cavity within the bony labyrinth. It is situated behind the cochlea and in front of the semicircular canals. Within the perilymph of the vestibule, there are two connected sacs of the membranous labyrinth, linked by a duct.

*The cochlea* is a conical coiled structure within the bone, approximately the size of half a pea. It makes about 2.75 turns and contains the cochlear duct, which houses the spiral or Corti's organ - the sensory organ of hearing. The cochlear duct, along with the spiral bony lamina, divides the cavity of the bony cochlea into three sections or scalae: the vestibular scalae, located above the cochlear duct, connected to the vesti-bule and pressing against the oval window; the middle scalae, which is the cochlear duct itself; and the tympanic scalae, located below the cochlear duct and ending at the round window. The cochlear duct, as part of the membranous labyrinth, is filled with endolymph. The vestibular and tympanic scalae, as part of the bony labyrinth, are filled with perilymph. At the apex of the cochlear spiral, they are interconnected through an opening called the helicotrema. The upper wall of the cochlear duct is formed by the vestibular membrane or Reissner's membrane, which separates it from the vestibular scalae. The outer edge of the vestibular membrane consists of a highly vascularized mucous membrane called the vascular stripe, which contributes to the formation of endolymph. The lower wall of the cochlear duct is formed by the spiral bony lamina and the flexible fibrous basilar or ground membrane, upon which the organ of Corti is situated. The basilar membrane plays a crucial role in sound perception; it is narrow and thick near the oval window and becomes wider and thinner towards the apex of the cochlea [2].

## 2. Existing Theories of Human Audition

There are various theories of audition that explain the mechanism of sound perception in the spiral organ, the receptor of the auditory system. We will also borrow these theories from open sources.

The theories of peripheral sound analysis suggest the possibility of primary analysis based on its properties in the cochlea, thanks to its anatomical and functional characteristics. Helmholtz's resonator theory posits that the basilar membrane is a set of "strings" of varying length and tension, similar to a musical instrument (e.g., a piano).

These "strings" resonate and respond to the corresponding frequencies of sound waves, similar to an open piano. Helmholtz's theory is supported by the morphological structure of the basilar membrane: at the base of the cochlea, the strings are shorter (0.16 mm), resonating with high-pitched sounds, while at the apex, they are longer (0.52 mm) and respond to low-frequency signals. When complex sounds are presented, multiple sections of the special membrane vibrate simultaneously, explaining timbre. The perception of sound intensity depends on the amplitude of membrane vibrations. Helmholtz's theory was the first to explain the fundamental properties of the ear, such as pitch, intensity, and timbre perception. However, it does not explain the phenomenon of masking, where high-frequency sounds are masked or obscured by low-frequency sounds. Furthermore, modern knowledge does not support the idea of individual "strings" on the basilar membrane, especially given the vast number required on a 35 mm membrane to perceive frequencies in the range of 0.2-20 kHz.

According to the hydrodynamic theory proposed by Békésy, when a sound wave travels through the perilymph of both scales, it causes oscillations of the basilar membrane like a traveling wave. Depending on the frequency of the sound, maximum curving of the membrane occurs at a specific point. Low-frequency sounds generate a wave that travels along the entire length of the basilar membrane, causing maximum displacement near the apex of the cochlea. Mid-frequency tones result in maximum displacement around the middle of the basilar membrane, while high-frequency sounds displace the basal turn of the spiral organ, where the basilar membrane is more elastic and flexible.

The Roaf-Fletcher hydrodynamic theory, based on Lutz's experiments with U-shaped tubes and liquid, supports Békésy's conclusions that high-frequency sound waves propagate near the basal coil of the cochlea, while low-frequency ones travel towards the helicotrema.

Flock (1977) suggests that in the formation of frequency selectivity, the main role is played by the basilar membrane with its outer hair cells, rather than the inner hair cells, as believed by many authors. These cells have efferent connections. The cilia are arranged in a rigid W-shaped structure, so any changes in the length of the cell due to potential differences will result in a shift of the basilar membrane. Actin and myosin have been found in the structure of outer hair cells, which are necessary components of any contractile system. The bioelectric activity of outer hair cells in the mechanical oscillation of the basilar membrane has been confirmed by studies (W. Brownell, G. Bander et al. 1985). Currently, there are mathematical and physical models of the hydrodynamics of the cochlea that include both nonlinear and active mechanisms (Shuplyakov B.C. et al. 1987; Zwicker E., 1986).

Ukhtomsky's theory of "physiological cell resonance" suggests unequal physiological lability of hair cells, which selectively respond to different frequencies of sound waves. With high lability of hair cells, they respond to high-frequency sounds, and vice versa, resembling physiological resonance.

The central theories of Rutherford and Ewald, unlike the previous ones, deny the possibility of primary sound analysis in the cochlea. According to Rutherford's telephone theory, the transmitting mechanism for all frequencies is Corti's membrane, resembling a telephone diaphragm with a microphone effect. When pressure is applied to the hair cells, the membrane transmits microphone-like potentials as synchronous signals to the brain centers, where they are analyzed. The role of mechanical vibrations of the basilar membrane is ignored in this theory.

According to Ewald's theory, sound causes "standing" waves on the basilar membrane, similar to Chladni patterns (sound images), which are then analyzed in the brain centers, resulting in various auditory sensations.

The dualistic theory by Riebel attempts to combine the spatial theory with the telephone theory. According to his theory, low-frequency sounds are transmitted directly to higher auditory centers, while high-frequency sounds have their precise localization in specific regions of the basilar membrane. This contradicts the facts because the impulses from higher CNS regions do not correspond to the frequency and character of the sound wave. For example, the frequency at the round window of the cochlea is 16,000 Hz, the auditory nerve is 3,500 Hz, the brainstem is 2,500 Hz, and the auditory cortex of the brain is 100 Hz.

*Otoacoustic emission.* In the oscillatory process and deflection of the basilar membrane, otoacoustic emissions (Kemp D., 1978; Kemp D., Chum R., 1980) may also play a role. It involves the generation of acoustic signals in the cochlea without auditory stimulation or after it, which are recorded using a miniature and highly sensitive low-noise microphone placed in the external auditory canal. These signals can vary in frequency and waveform among different individuals. The individual pattern of emissions may correspond to individual deviations in the audiogram. In cases of inner ear pathology, the emissions' "thresholds" may change. Spontaneous emissions that occur without acoustic stimulation are attributed to the role of sharp frequency tuning of the cochlea and the activity of auditory receptors. Perhaps there is a simultaneous influence of the traveling wave and resonance on the basilar membrane due to otoacoustic emissions.

*Contemporary Understanding:* Previous concepts of the inner ear structures and their functions, upon which biophysical models of hearing were based, have serious limitations. These concepts are overly simplified, often depicting the cochlea as consisting of just two similar chambers (spaces, cavities, or canals) filled with perilymph and separated by a so-called equivalent membrane. The cochlear models do not take into account either the anatomical peculiarities of the inner ear structures or the physiological (functional) properties of the fluids filling the corresponding spaces of the inner ear and the membranes that separate them, nor the physical properties of the inner ear structures. Therefore, they do not elucidate the real biophysical processes underlying the nature of hearing.

However, based on these cochlear models, two fundamentally different models of auditory perception have been created, along with a significant number of their variations and combinations. One of the earliest models of hearing was proposed by H.L.F. Helmholtz [10] (explained in [11, 12]). According to H. Helmholtz, the inner ear is considered to consist of two chambers filled with perilymph and divided along the cochlea by the basilar membrane, which determines the nature of hearing. According to H. Helmholtz, mechanical vibrations of the membrane of the oval window are transmitted to the perilymph, which, in turn, causes vibrations of the basilar membrane. This modern model, which fairly comprehensively describes many phenomena, formed the basis for experiments by G. Bekesy and subsequent biophysical developments, the full review of which is presented by V.S. Shuplyakov. H. Helmholtz's mechanism of hearing, and accordingly the mechanism based on it, is still presented practically without variations in all existing textbooks on physiology, anatomy, and histology.

The description of the hearing mechanism according to H. Helmholtz still has several difficulties [11, 12]. One of them is that the basilar membrane must be constantly in a state of tension, which is not observed in experiments [13]. Another problem with the theory of hearing by H. Helmholtz is that it cannot explain the perception of a "missing fundamental" frequency at 100 Hz, which is physically absent when there are tones at 1100, 1200, and 1300 Hz.

The most well-known alternative model of hearing is the so-called telephone theory proposed by W. Rutherford [23] (explained in [12]). According to W. Rutherford, the cochlea is not frequency-dependent along its length, and all its segments respond to all frequencies simultaneously. The cochlear mechanism transmits all the wave parameters of the sound signal to the central nervous system (CNS), where sound analysis takes place.

This theory also faces difficulties in explaining the mechanism of peripheral auditory reception. The first problem is the impossibility of transmitting a signal with a frequency above 1 kHz due to the presence of an absolute refractory period in nerve fibers (with a duration of about 1 ms). This problem was positively addressed by the introduction of the volley principle, proposed by E.G. Wever [24]. Another complexity of the model lies in its inability to explain the "apical-basal effect," which involves the loss of the ear's ability to perceive high-frequency components of a wave signal when there are disruptions in the apical portion of the basilar membrane.

This bio-physical and wave model of human hearing deserves attention for its method of modeling the biophysical processes that realize the mechanism of hearing in humans. This model includes identifying the anatomical and histological structures in relation to the distribution of sound velocity by frequencies; formation of standing waves on the basilar membrane (Reissner's membrane) as a result of interference between waves traveling along it and waves reflected from its free edge near the helicotrema; establishing the relationship between the distribution of sound velocity and the coordinates of the basilar plate by frequencies, with a focus on the point of maximum mechanical impact on the nerve endings generating nerve impulses in the hair cells, corresponding to the frequency of the outgoing sound signal [25]. All other models of the ear and mechanisms explaining the nature of auditory reception are combinations of the above.

### 3. Drawbacks of Existing Human Auditory Systems

Therefore, despite a large number of studies, we currently do not have an adequate model of the human auditory system that would correspond to the physical processes occurring in the human ear in full accordance with its anatomical structures. So, let's formulate the drawbacks of existing auditory models in the form of a series of questions for which there are no answers in the relevant literature.

- What role does the external part of the ear play, and why does it have such a structure?
- The tympanic membrane has a logarithmic relationship, but it is known that the logarithm of negative numbers does not exist, so how is logarithming accomplished?
- What role does the middle ear cavity play?
- What role does the system of the malleus, incus, and stapes play, and if it provides amplification, why doesn't it increase the amplitude of vibrations?
- What type of wave travels within the cochlea of the ear?
- Why does the inner surface of the cochlear spiral have ridges or "hair cells"?
- Why is the cochlea coiled into a spiral shape?
- What role does the Reissner's membrane play when it is often overlooked in models?
- What role does the helicotrema play?

- How is the analysis of frequency oscillations carried out when it is known that they are converted from time to distance, meaning they are location-dependent?
- How is the fast action of ear neurons synchronized with the frequencies the ear perceives (the firing rate is known, which is 1 kHz, but we hear 10 kHz), given that Shannon's theorem doesn't apply?
- Where does the mel scale, used in speech recognition, come from?
- What role does the auditory (Eustachian) tube play?
- How do we reproduce speech where the parameters of the cochlea match the parameters of the human vocal tract?

Can the resonance of the basilar membrane, when considered as tensioned strings of a piano, provide selectivity (selection) of sounds, such as musical ones, for which this selectivity is equal to:

$$\left.(f + f * \sqrt[12]{2})\middle/(f - f * \sqrt[12]{2})\right. = \left.(1 + \sqrt[12]{2})\middle/(1 - \sqrt[12]{2})\right. \cong 36\,, \qquad (3.1)$$

and what do we clearly distinguish?

So, on these and other questions, we do not find answers in existing models of the human auditory system. This is because in these theories, each part of the ear is considered separately from the others, which does not allow for a comprehensive understanding of the functioning of the human auditory system.

## 4. The Cochlea and the Traveling Wave

Let's consider the main organ of hearing - the cochlea (snail) in Figure 2. In this illustration, on the left, the cochlea is schematically depicted. In reality, humans do not have a separate organ like this. This is for illustration purposes only. Instead, there are only channels (labyrinths) within the solid temporal bone. In reality, it is as shown on the right in the figure, where a cross-section of this bone is depicted. The cochlear canal or bony labyrinth is a flattened cone. The bony labyrinth is a dense bony structure, the only bone structure in the body where bone remodeling does not cease. In the cochlea, the bony part is represented by the cochlear spindle and the cochlear spiral canal, which surrounds the spindle 2.5 times. From the spindle, a bony spiral plate extends, which, together with the basilar membrane of the cochlear spiral, divides the lumen of the canal into the vestibular stairs connected to the oval window, and, together with the Reissner's membrane of the cochlear spiral, into the tympanic stairs closed by the secondary membrane of the round window (Figure 2). The tympanic and vestibular stairs are filled with a fluid called perilymph. They connect at the apex of the cochlea through the helicotrema. The membranous part of the cochlea forms the cochlear spiral, which has a triangular shape in cross-section, formed by the aforementioned membranes: the basilar membrane at the bottom and Reissner's membrane at the top. The cochlear spiral, located between the vestibular and tympanic stairs, forms the so-called middle stairs, filled with endolymph. It terminates blindly at both ends: at the top, adjacent to the helicotrema, and at the bottom, at the vestibular side.

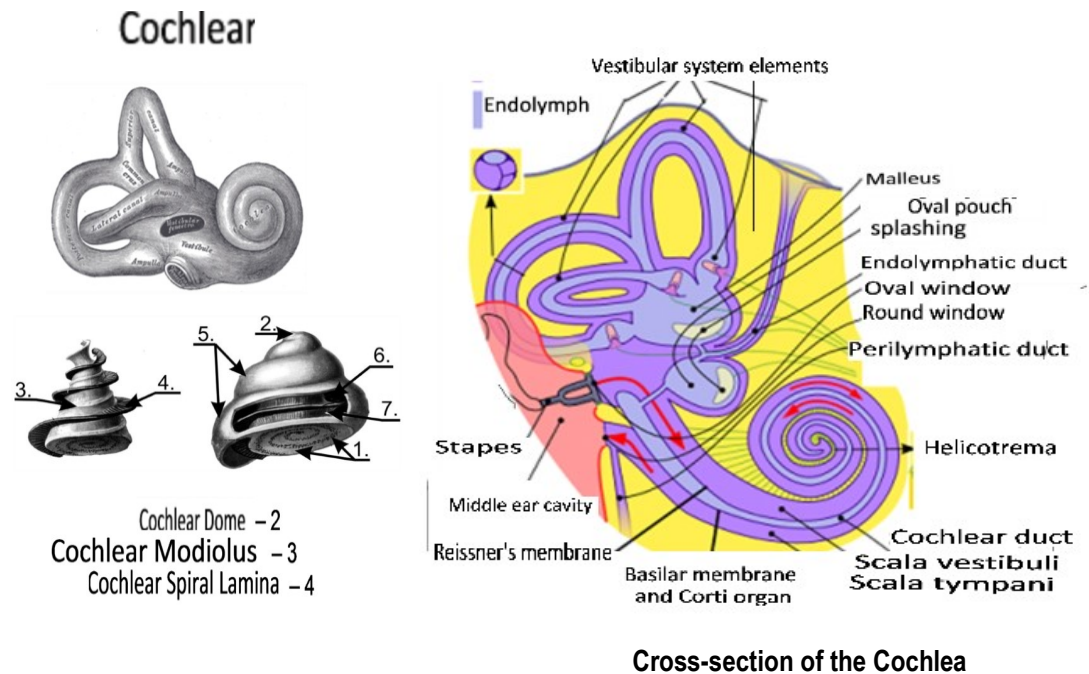

Cross-section of the Cochlea

Figure 2 – The cochlea of the inner ear.

Note that for sound waves in air, the speed of sound is 343 m/s (under normal temperature and atmospheric pressure conditions). The human ear can perceive sound wave lengths corresponding to sound frequencies ranging from 20 Hz to 20 kHz, which would fall within the range of approximately 17 meters to 17 millimeters, respectively. However, the cochlear canals are filled with fluid where the speed of sound is even greater (1500 m/s). This would indeed result in the wavelengths being increased by a factor of three. Yet, the length of the cochlea is only about 3.5 cm. This raises the question: "What type of wave is actually propagating there?"

Typically, in most literary sources, the functioning of the cochlea is illustrated with Figure 3 https://studfile.net/preview/5600004/page:6/. As seen in this figure, the stapes vibrates the fluid in the vestibular canals. From Wikipedia https://uk.wikipedia.org/wiki/Вухо.

Sound waves enter the external auditory canal and cause the eardrum to vibrate with the corresponding frequency and amplitude. The greater the intensity (loudness) of the sound, the greater the amplitude of the eardrum's vibrations. The motion of the eardrum is then transmitted to the auditory ossicles, which act as levers and set the oval window of the inner ear in motion. Because the area of this window is 20-22 times smaller than the eardrum, the amplitude of vibrations increases here by a corresponding amount, leading to signal amplification. Oscillating back and forth at a certain frequency, the oval window of the inner ear causes a similar movement of perilymph in the vestibular canal, which extends from the base to the apex of the cochlea. Sound at very low frequencies (below 16 Hz) creates pressure waves in the fluid that travel the entire path from the oval window, up the vestibular canals, through the helicotrema to the tympanic canals, and finally transmit to the round window of the inner ear. Such sounds do not stimulate the corti organ and are beyond the range of human hearing (infrasound). Sound waves at higher frequencies create pressure waves that can "cut" their way through the vestibular canal to the perilymph of the tympanic canals. In this case, they induce vibrations in the basilar membrane, causing the covering membrane to shift, rhythmically stimulating the hair cells.

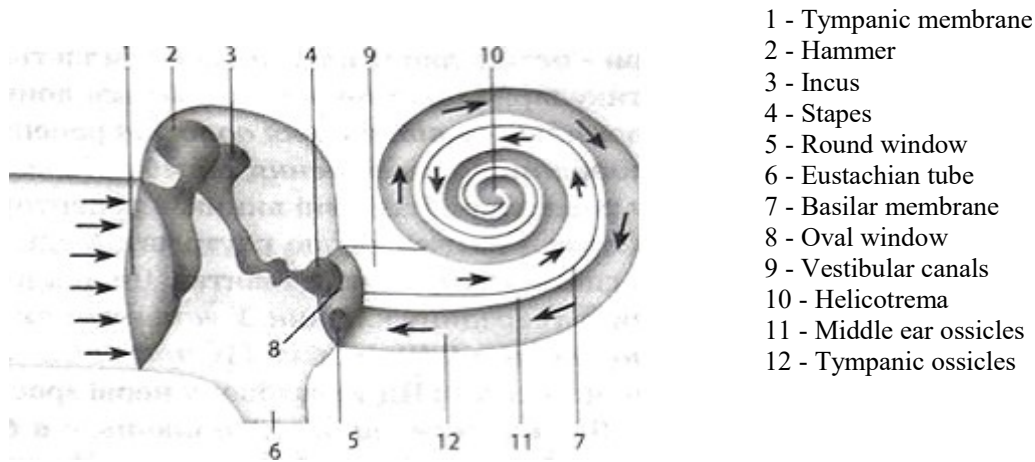

1 - Tympanic membrane
2 - Hammer
3 - Incus
4 - Stapes
5 - Round window
6 - Eustachian tube
7 - Basilar membrane
8 - Oval window
9 - Vestibular canals
10 - Helicotrema
11 - Middle ear ossicles
12 - Tympanic ossicles

Figure 3 - Pathways of sound wave propagation in the cochlea.

But this is complete nonsense! It's also strange that textbooks write this way. If we assume, according to this theory, that the tympanic membrane has deflected at least 1 mm, then the stapes should deviate by 2 cm, which is more than half the length of the cochlea. But what is the reality? The pressure of sound on the tympanic membrane is insignificant. To disturb the viscous perilymph, you need to increase the force, which means pressure. However, the laws of mechanics, including fluid mechanics, say that to increase force, you need to win in distance. In other words, increasing the amplitude by 20-22 times is not possible, but increasing the pressure is. Because pressure is force per unit area. Since the area of the stapes is smaller than the area of the tympanic membrane, the pressure of the stapes (force per unit area) on the perilymph is greater and sufficient to disturb this fluid. That's the first point. Second, why would the oscillation of the fluid occur in a large circle from the stapes to the helicotrema, and then through the helicotrema, along the cochlear ducts, to the round window, when there is a shorter path (see Figure 4).

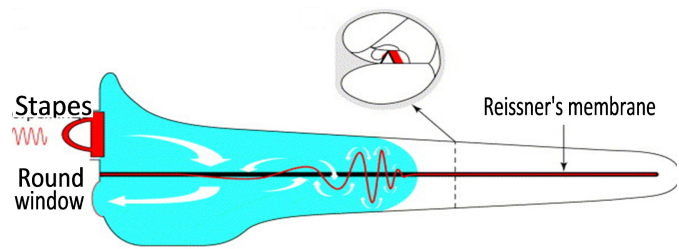

Figure 4 – Oscillation of perilymph in the cochlea

Therefore, when the stapes vibrates, a reciprocating flow of fluid will occur between the stapes and the oval window. If the stapes moves inside the cochlea, it will push the oval window outward, and vice versa. This creates a fluid relaxator in the vestibule of the cochlea that will replicate the forced vibrations of the stapes. Since there are two membranes in the path of this flow, the Reissner's membrane and the basilar membrane, a small portion of these membranes in the vestibular region will also vibrate synchronously under the influence of this flow. Further into the cochlear ducts, the fluid will not move but will only change pressure. Due to the change in pressure in the Reissner's membrane, a well-known traveling wave will occur, as shown in Figure 4. This is the same kind of wave that occurs in our blood vessels when the heart contracts and ejects a portion of blood. The conditions for its formation are just ideal: on one side, there is the bony surface of the vestibular canals, and on the other side, there is the stretched, homogeneous, thin, and elastic Reissner's membrane. The existence of waves in the Reissner's membrane has been confirmed in a study [26], from which Figure 5 is borrowed.

However, in the basilar membrane, a pulsatile wave will not form because this membrane is amorphous and not stretched, being anchored on both sides. But the oscillations created by the pulsatile wave of the Reissner's membrane will be transmitted through the middle channel filled with endolymph and reach the basilar membrane, as shown in Figure 5D. This means that the basilar membrane will simply replicate the oscillations of the Reissner's membrane. Further, the oscillations caused by the Reissner's membrane will propagate along the vestibular canal of the cochlea to the helicotrema in the form of a traveling wave. Part of the wave reflects off the helicotrema, and part of it passes through.

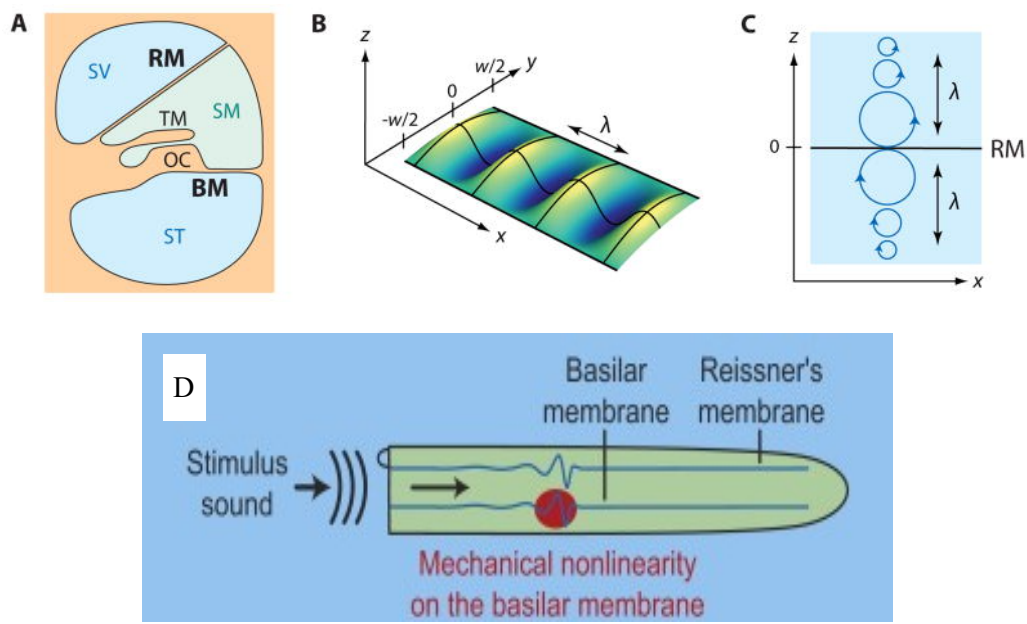

Figure 5 – Mechanism of generating a traveling (pulsatile) wave in the Reissner's membrane.

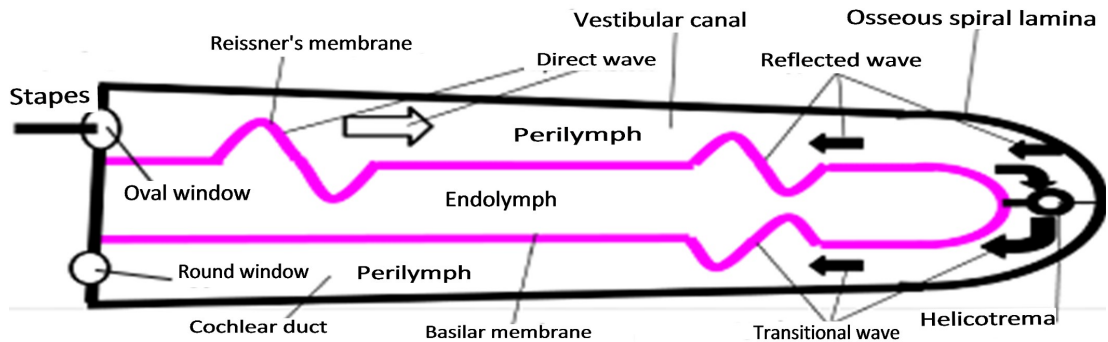

Figure 6 – Compensation of the reflected wave.

The reflected wave changes its phase by 180 degrees, while the wave passing through the helicotrema does not change its phase. Thus, the reflected wave in the vestibular scala and the direct wave in the tympanic scala compensate each other in the middle scalae of the cochlear duct between the Reissner's and basilar membranes. Therefore, there is no reflected wave in the cochlear duct, as shown in Figure 6.

Notice that the velocity of the traveling wave is almost two orders of magnitude lower than the velocity of sound and is determined by the Moens-Korteweg equation:

$$u = \sqrt{\frac{hE}{2\rho R}}$$

(4.1)

where: $E$ is the Young's modulus, $h$ is the thickness of the basilar membrane, $R$ is the radius of the basilar membrane, $\rho$ is the density of the fluid. For an artery, this velocity ranges from 5 to 10 meters per second. In our case, this will be 2-4 meters per second, which is proportional to the dimensions of the cochlea.

## 5. The Phenomenon of Transverse Resonance

As mentioned earlier, a separate group of theories in the field of hearing comprises the "central analyzer" theories, often referred to as telephone theories. According to these theories, sound vibrations are transformed by the cochlea into synchronous waves in the nerve and transmitted to the brain, where their analysis and pitch perception occur. This group of theories includes I. Evald's theory [27], according to which, when sound acts on the cochlea, standing waves are formed with a length determined by the frequency of the sound. The perception of pitch is determined by the perception of the pattern of these standing waves. The sensation of a particular tone corresponds to the excitation of one part of the nerve fibers, while the sensation of another tone corresponds to the excitation of another part. Sound analysis does not occur in the cochlea but in the centers of the brain. Evald managed to construct a model of the basilar membrane, the size of which roughly corresponds to its real size. When it is excited by sound, the entire membrane vibrates in a oscillatory motion, and a "sound picture" in the form of standing waves called "Chladni figures" is created, with a length shorter than the higher-pitched sound. Despite the good explanation for some complexities, Evald's theory, like other "central analyzer" theories, poorly aligns with recent physiological research on the nature of nerve impulses. Therefore, some researchers consider it possible to adopt a dual perspective, explaining the perception of high-pitched tones (which do not encounter complications) from the perspective of the "peripheral analyzer" theory and low-pitched tones from the standpoint of the "central analyzer."

Indeed, according to this theory, all information about the frequency and intensity of acoustic oscillations, like in the telephone theory, falls within the realm of higher structures in the auditory system, which has not been confirmed further. As a result, Evald's theory did not gain significant importance and today holds only historical interest.

That's a pity. Let's now consider a scientific work titled "Transverse Acoustic Waves in Rigid Tubes"[26]. Below, we provide an abstract by the authors of this work. Recently, significant work has been conducted, both theoretical and experimental, regarding the transmission of electromagnetic waves through dielectric waveguides and hollow metallic tubes. A similar phenomenon is the propagation of transverse acoustic waves in gaseous media confined by rigid cylindrical tubes. The paper describes the apparatus and methods for generating certain lower-order acoustic waves, and measurements are performed using a circular cross-section tube filled with air. The measured values of wavelength, frequency, and pressure distribution for certain possible types of harmonic waves were found to be in excellent agreement with theoretical values. A measurement of absorption using a standing wave method with transverse waves is proposed.

This publication reveals that, contrary to the commonly held belief of the existence of only longitudinal standing waves in a cylindrical tube, the experiments conducted by the authors have demonstrated the existence of transverse standing waves as well. This is a crucial factor for our research.

Moreover, in the work [28], which is dedicated to multi-wave waveguides with random irregularities, the phenomenon of transverse resonance is observed and described. The essence of the transverse resonance phenomenon lies in the fact that when a traveling wave propagates in a waveguide, due to the non-uniformity of the inner surface of this waveguide, wave diffraction occurs. As a result, a standing wave is created in the transverse direction to the wave propagation, if the transverse diameter of the waveguide is a multiple of half the wavelength, regardless of the fact that the wave is longitudinal. Subsequently, the standing wave accumulates the energy of the traveling wave, significantly increasing its amplitude, which is precisely the phenomenon of transverse resonance.

In our case, the transverse component of the traveling wave, generated by the Reissner's membrane, passes through the basilar membrane into the cavity of the tympanic ducts within the cochlea's middle turns. There, the distance between the basilar membrane and the inner bony surface decreases due to the deformation of the round window membrane, as perilymph, being a liquid, is incompressible but viscous. Then, this wave reflects from the bony surface of the tympanic ducts and travels back through the basilar membrane, passing through the cavity of the middle turns, and returns through the Reissner's membrane into the cavity of the vestibular ducts. In the location where the cross-sectional area of the cochlear duct (its diameter) equals half the wavelength, a standing wave is formed, as shown in Figure 6, because there will always be a corresponding cross-sectional area in the tapered cochlear duct with a diameter D=λ/2. In this case, we are dealing with a transverse standing wave. Since the cochlea has a conical shape, in the cross-section, this wave is approximately circular and has a pattern as shown in Figure 7.

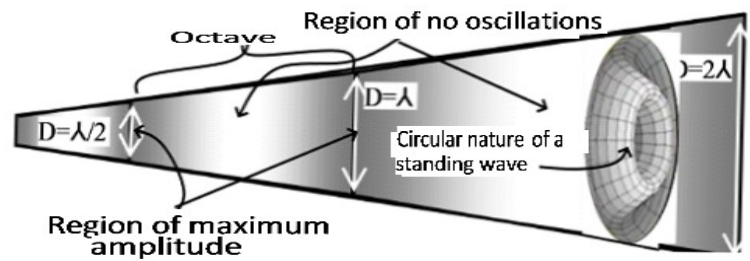

Figure 7 - Pattern of a fixed-frequency standing wave.

Let the equation of the direct transverse wave be s1 = A * cos($\phi$), and the reflected wave be s2 = A * cos($\phi$ - $\phi$(D)), where $\phi$ is the phase of the direct transverse wave, and $\phi$(D) is the phase of the reflected wave, which is a function of the diameter of the cross-section of the cochlear duct, and A is its amplitude. Then their sum is given by:

$$S = s_1 + s_2 = 2A \cdot \cos(\phi - \phi(D)/2) \cdot \cos(\omega t + (\phi + \phi(D))/2) \qquad (1)$$

From this expression, it can be seen that the term $2A \cdot \cos(\phi - \phi(D)/2)$ represents the amplitude of this oscillation, while $\cos(+(\phi + \phi(D)/2)$ describes the oscillatory process itself. Obviously, the phase difference $(\phi - \phi(D)/2)$ depending on the diameter of the cochlear duct will be exactly zero in this place because the amplitude in this place is the largest and equal to 2A.

It is also clear that moving away from this point, this amplitude will decrease according to the cosine law, and in the cross-section of the cochlear duct where the phase difference becomes exactly 90 degrees, the amplitude will be zero. Moving further away from this point, we will observe an increase in the amplitude of oscillations again. This amplitude will reach its maximum again in the cross-section where it is a multiple of D= 2*λ/2, and so on. It is evident that we are dealing with wave interference [3], i.e., the alternation of maxima and minima, as shown in Figure 7. This alternation in the form of the amplitude graph of the Reissner's membrane oscillations is shown in Figure 8.

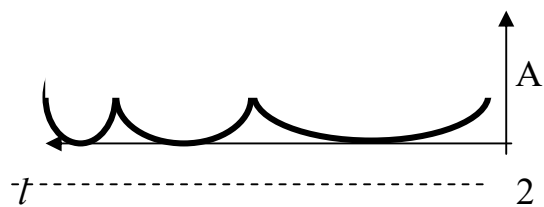

Figure 8 – Dependence of the amplitude of a fixed-frequency standing wave on the distance in the cochlea (A – pressure, l – distance in the cochlea).

However, humans can distinguish musical compositions quite well, meaning that the quality and selectivity (the ability to discriminate) of the human auditory system are relatively high and determined by the relationship described in equation (1). Passive systems, such as the human ear, are passive because they do not contain amplifying elements and have a quality factor (Q-factor) of around 1-2, as shown in figure 8. However, there is also the phenomenon of transverse resonance, which allows achieving such high quality. Irregularities or inhomogeneities in the waveguide's inner surface (figure 9) contribute to the occurrence of this transverse resonance. Since sound signals over a short time interval represent a specific frequency, the standing wave behaves like a "black hole" that absorbs the energy of individual harmonics of the traveling wave until equilibrium is reached between the energy influx from the traveling wave and the losses of this energy due to internal friction in the perilymph, as in a viscous fluid. Then, the amplitude of the standing wave increases significantly and can be precisely processed by a threshold element (neuron), which is essentially the phenomenon of transverse resonance. In waveguides, this is a parasitic phenomenon that disrupts their normal operation, but evolution has turned it into a useful phenomenon. Thus, this process can be likened to the operation of a well-known rejector filter, which absorbs energy at a specific frequency without affecting others and is widely used in radio engineering.

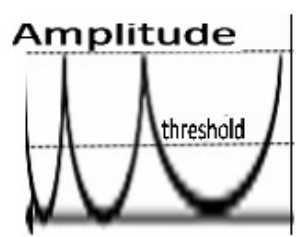

Figure 9 - Notches on the inner surface of the cochlea

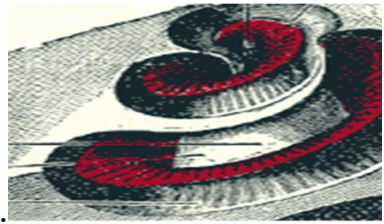

**Рис. 10** – The phenomenon of transverse resonance of a fixed frequency

## 6. Sound Processing Principle in the Cochlea

Up to this point, the shape of the cochlea had not been taken into account. It has a conical shape. So, by choosing two closely spaced diameters of the cross-section of the cochlear duct, which is conical in shape, in a ratio of one to two and dividing this section of the duct into 12 equal parts, we obtain standing waves at these intersections with lengths corresponding to the frequencies of the musical scale, i.e. $f_{i+1}/f_i = \sqrt[12]{2}$ .

The number of frequencies in this section is an octave, as shown in figure 11. Moreover, this distribution, which is spread along the length of the cochlea and is tied, so to speak, to a specific location within this distribution, is almost linear (the error fluctuates in hundredths of a millimeter). Therefore, it is the conical shape of the cochlea that allows for the analysis of the frequency composition of sound signals.

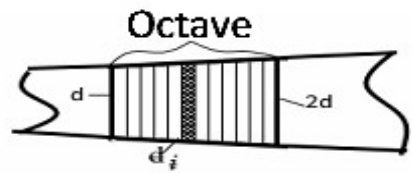

Figure 11 - Distribution of wavelengths (frequencies) in the cochlear duct

$$(di=\lambda i/2=d \cdot \sqrt[\tfrac{1}{2}]{2})$$

Most models of the ear consider the cochlea as unrolled, assuming that the coiled spiral shape (as shown in Figure 2) does not affect the essence of the physical processes occurring within the inner ear [5]. However, this is not the case. Based on research on otoacoustic emissions conducted by researchers at Rockefeller University [4], the wave generated in the Reissner's membrane is a spherical wave, as shown in Figure 4-B. This fundamentally changes the situation. A spherical wave is a wave whose front is a sphere. The phase velocity vector of a spherical wave radiates and is oriented radially from the source (the wave either radiates radially from the source or converges towards it). Then, in the transverse cross-section of the cochlear spiral, we will have a circular wave limited by the inner surface of the cochlear duct, which can be illustrated as shown in Figure 12 (sourced from the internet). https://www.slideshare.net/AlexVoron-kin/3-10848108). Clearly, the energy of the circular wave, which is the intersection of the spherical wave, will retain its value at this point, except for losses due to friction in the fluid.

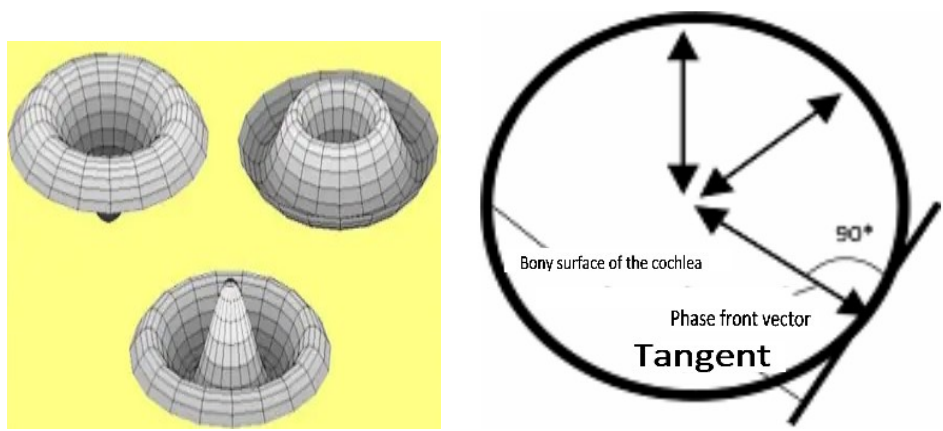

Figure 12 – Circular wave in the cross-section of a cochlear turn.

However, if the cochlear spiral had an unfolded form, there would be no standing spherical wave because its energy would spread along the entire length of the spiral (Figure 13a). Evolution, however, solved this problem by twisting the cochlear cone into a spiral in such a way that the angle of incidence of the wave would equal the angle of the reflected wave, allowing for the focusing of the energy of the spherical wave (Figure 13b).

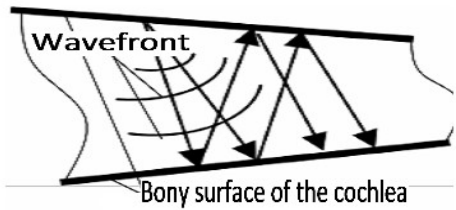

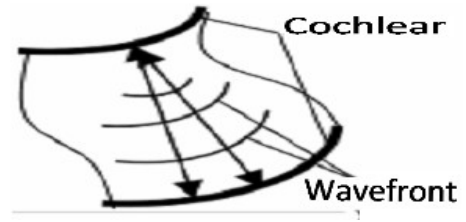

Figure 13a – Unfolded cochlear spiral       Figure 13b – Spirally coiled

Therefore, the cochlea has a conical spiral shape (Figure 14a), and Figure 14 shows the mechanism of exciting standing waves in the cochlea in a generalized form. It should be noted that the cochlea is coiled in a specific way. The reason is that according to Figure 13b, it can be seen that the energy of the wave is focused only on the reverse path of the standing wave. However, the cochlea has a cross-section shown in Figure 14b and Figure 2. The wave oscillates along the outer wall of the scala vestibuli and the scala tympani (Figure 14b), and in this case, the wave is focused both on the direct and reverse paths of the wave.

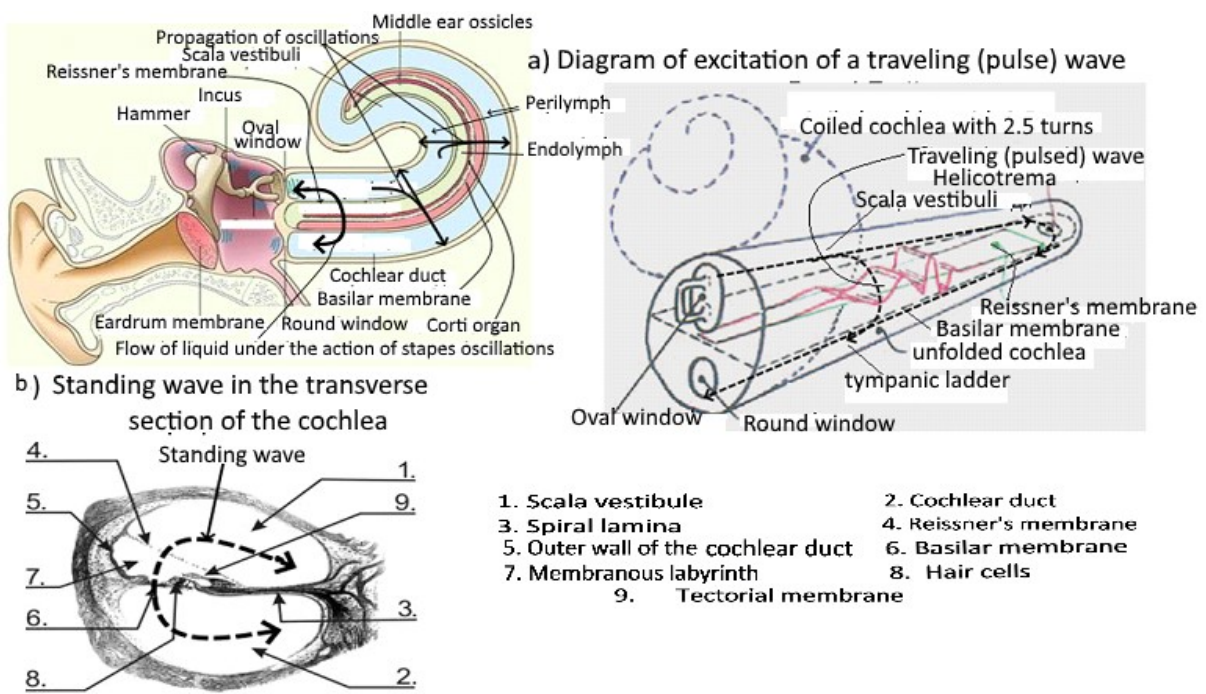

Figure 14 – Mechanism of generating standing waves in the cochlea

It's worth noting that at the points of maximum amplitude where the nodes of the standing wave are located, they will be on the bony surface of the cochlear duct. In the transitional zone, we can't make the same statement because there's no information about the velocity of the traveling wave in the Reissner's membrane. However, this is not that important; what matters is that in the points of minimum and maximum amplitude, a stable process will occur, and these points will be anchored, so to speak, to their location. Then, two membranes (Reissner's membrane and the basilar membrane) will oscillate synchronously at these points, but with different phases, but with a constant phase difference between them. A constant phase difference at the points of maximum and minimum (in the standing wave region) will create an almost constant pressure between these membranes in the scala media (as shown in Figure 15).

This will, in turn, lead to an almost constant (during the oscillation) partial volume of fluid (endolymph) between these membranes and, as a result, a constant (during the oscillation) charge in this section of the fluid. This charge is detected by the hair cells of the basilar membrane and is further transmitted to the neurons of the auditory system.

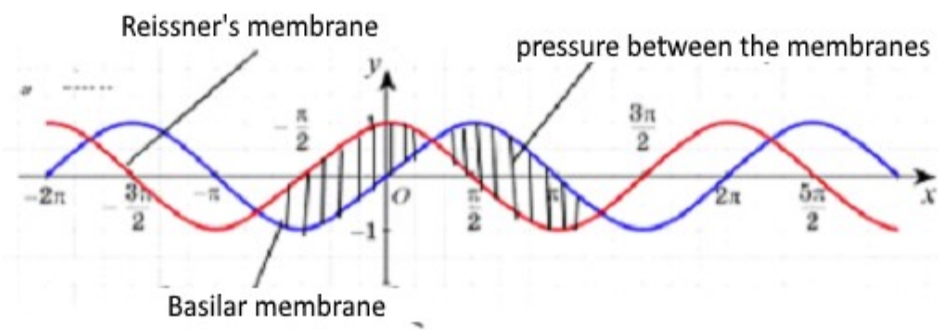

Figure 15 – Pressure relationship between the Reissner's membrane and the basilar membrane

Since the duration of sound signals perceived by living organisms in time, ranging from 20 ms for speech signals to 200 ms for musical sounds, is significantly longer than the periods of its component frequencies, and the pressure difference between the membranes will remain almost constant during the time of action of a particular sound signal, the neural network of the brain will sample their characteristics at least 20 times within such a time frame, assuming the fast firing rate of neurons at 1 kHz. Thus, in technical terms, nature or evolution has created, based on standing waves, a well-known sampling and storage device, akin to those used in computing technology. Also, it's worth noting that the mentioned interference cannot be captured by a high-speed camera, as we can do for light waves because it develops over time and is directly related to the speed of the traveling wave in the Reissner's membrane. In other words, the peaks of amplitude values of standing waves, as discussed earlier, are not simultaneous but are functions of time. On the other hand, the alternation of these peaks resembles the process of running lights, creating the illusion of a traveling wave in the basilar membrane, as determined by Bekesy's experiments. Therefore, there is no traveling wave in the basilar membrane in its classical understanding, but its illusion is created by the interaction of transverse standing waves in the cochlear duct.

However, the question arises of how we determine the frequencies of the sounds that enter our ears when we are not anchored to any known frequency. But this is not important. Humans perceive not the absolute value of frequency but its relative value.

This is why we are not surprised when we listen to a melody that starts in one key or another. Modern musical instruments, with their tempered tuning, allow us to start a melody on any note, and we still understand that melody. Similarly, we understand language, regardless of who is speaking (a man, a woman, or a child). In other words, the relative perception of sound frequency is a normal phenomenon, and the key criterion is not the absolute value of this resonant frequency of the middle ear but its stability over a significant period of time.

And how is this frequency perceived by the ear? Let's turn to the amplitude response of the tympanic membrane. Figure 16 shows this characteristic as a function of the signal amplitude [29] coming from the surrounding environment.

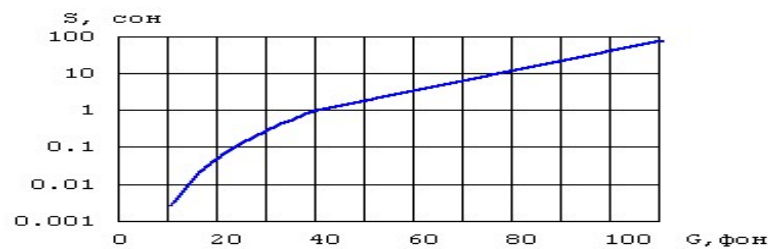

Figure 16 - Dependence between loudness and the level of the tympanic membrane.

From this figure, it can be seen that this is a logarithmic relationship, and its form is associated with the need to reduce the dynamic range of the input signal and protect the auditory system from overload, as excessive signal levels can lead to hearing loss. On the other hand, from the theory of measurements, it is known that when you need to accurately measure frequency, you need to stabilize the amplitude of the signal. Conversely, when you need to accurately measure the amplitude of this signal, you need to stabilize its frequency, and all of this directly follows from the Heisenberg Uncertainty Principle. Therefore, the eardrum also performs this function. However, it has a nonlinear characteristic. From the course of radio engineering, it is known that if you apply two signals of different frequencies to a nonlinear element (in our case, a signal with the resonance frequency of the middle ear cavity and a signal from the surrounding environment), then, whether we like it or not, combination frequencies like fr-fs and fr+fs arise, where in our case, fr is the resonance frequency of the middle ear cavity, and fs is the frequency of the signal from the surrounding environment. Since the cochlea is a low-frequency analyzer, it perceives only the low-frequency component, i.e., fr-fs. From the perspective of radio engineering, we have a superheterodyne direct conversion receiver. It may seem like we invented this receiver, but as it turns out, evolution has been using it for a long time.

To confirm the analysis of the frequency difference by the cochlea, let's examine the frequency distribution along the basilar membrane taken using the "black box" principle [29] (an etalon signal of a predefined frequency is applied, and the response of the basilar membrane is recorded, Figure 17). From this Figure 17a, it can be seen that these frequencies are not evenly distributed: low frequencies are sparse, and high frequencies are dense. High frequencies are located in the wider part of the cochlea, while low frequencies are in the narrower part. However, this contradicts the phe-nomenon of transverse resonance and the theory of signals [30], which suggests that the higher the frequency of a signal, the wider the frequency band it occupies. In this case, we have the opposite situation. Such a distribution (Figure 17b) is interpreted as the basilar membrane having resonant properties (similar to the strings of a piano), and the selectivity of the system as a whole is based on these properties. Since these characteristics were measured without considering the frequency difference, we have a mistaken frequency distribution or an illusion associated with the "black box" method. If we take into account the frequency difference, we will have the following result. Since the cochlea perceives the frequency difference, it perceives a large frequency difference (high frequencies) in the narrow part of its channel and a small difference (low frequencies) in the wider part, in accordance with the phenomenon of transverse resonance.

If the frequency difference is small, it means that a high-frequency signal is applied to the input of the auditory system, and its response will be recorded in the wider part of the cochlea, i.e., at the entrance of the cochlea. If the frequency difference is large, it means that a low-frequency signal is applied to the input, and its response will be recorded in the narrower part of the channel. Therefore, in practice, the frequencies in Figures 17a and 17b should be swapped. Where 50 Hz is shown, 20 kHz should be placed, and vice versa. Then everything aligns with the theory of signals.

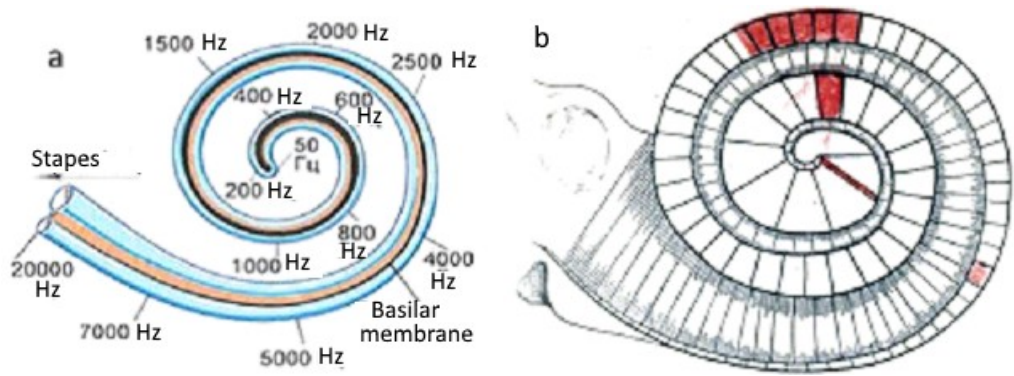

Figure 17 – Distribution of frequencies along the basilar membrane in existing theories of hearing.

It is also known that the basilar membrane has uneven elasticity and width. So at the entrance of the cochlea, where, according to our data, low-frequency signals are analyzed, it has greater rigidity, narrower width, and is thicker. At the apex of the cochlea, it is wider, thinner, and less rigid because high-frequency signals are analyzed there. However, such a structure is not dictated by resonance properties. Recall that Heisenberg's uncertainty principle states that it is impossible to simultaneously determine the amplitude of a signal and its frequency. Therefore, to determine the frequency of such a signal, it must be normalized, i.e., its amplitude must be equalized at a certain constant level. The tympanic membrane partially performs this function, but not at a sufficient level. Although logarithmic scaling narrows the dynamic range, a large amplitude will still be greater than a small amplitude. To compensate for this, evolution chose the basilar membrane with these properties (Figure 17b). If we also consider the damping of the traveling wave in the Reissner's membrane, such a shape of the basilar membrane ultimately evens out the unevenness of the acoustic amplitude-frequency characteristic of the cochlea.

## 7. The Role of the Middle Ear

Without the participation of the middle ear, it is impossible to fully understand the functioning of the ear. It plays a significant role in the human auditory system. Firstly, it deals with the noise of the surrounding environment. When a person is in absolute silence, it can lead to discomfort in their behavior, and here's the explanation for that. The noise from the surrounding environment, entering through the eardrum into the middle ear cavity, causes vibrations at a frequency of approximately 2 kHz because the middle ear cavity behaves like a resonator. This frequency leads to vibrations of the eardrum, which are transmitted through the ossicular chain to the stapes. During this process, the speed of the sound wave decreases, aligning it with the speed of the wave on the basilar membrane. Thus, the ossicular chain acts as an acoustic "speed transformer" for the sound wave. The resulting wave in the basilar membrane generates the pattern mentioned earlier (Figure 4). This pattern is the initial reference system relative to which the subsequent analysis of sound signals will be conducted. In other words, these vibrations calibrate our analyzer, which takes the form of a cochlea, and the specific dimensions and diameter of the cochlea hardly affect the quality of the analysis. What matters is that the cochlea has a conical shape. This approach allows for changes in the dimensions of the cochlea or its growth. It is known that the cochlea is the only bone system in humans that changes its size throughout life.

So, in absolute silence, a person, like other living beings, loses their auditory reference points, which can lead to discomfort. Next, let's assume that a sound signal of the form is input to our auditory system. Since the tympanic membrane logarithmically compresses the input signal (as shown in Figure 16), we will have:

$$\log y = \log[A\,cos\,(\omega t)] = \log A + \log\,cos\,(\omega t)$$

In this case, it should be noted that the tympanic membrane is stretched and displaced inside the middle ear due to the pressure difference between the surrounding environment and the middle ear cavity. Additionally, the ear canal, through which sound enters the ear, has a conical shape, leading to a deformation of the wave such that it is shifted towards positive amplitude values.

This ultimately leads to the fact that the expression for the input signal will have a constant component $y = A\,cos\,(\omega t) + c$ i, and then we have:

$$\log y = \log A + \log[cos\,(\omega t) + c], \tag{7.1}$$

which is shown in the graph below in Fig. 18.

From the graph, it can be seen that the function $\log[\,cos\,(\omega t) + 2]$ hardly differs from the function $cos\,(\omega t) + 2$. It maintains the positions of extrema relative to the abscissa axis, has reduced amplitude, and is likely not a tonal signal because it will have some high harmonics. Therefore, expression (7.1) can be rewritten as:

$$\log y = \log A + c + a_1 \, cos(\omega t), \tag{7.2}$$

where $a_1$ is a constant.

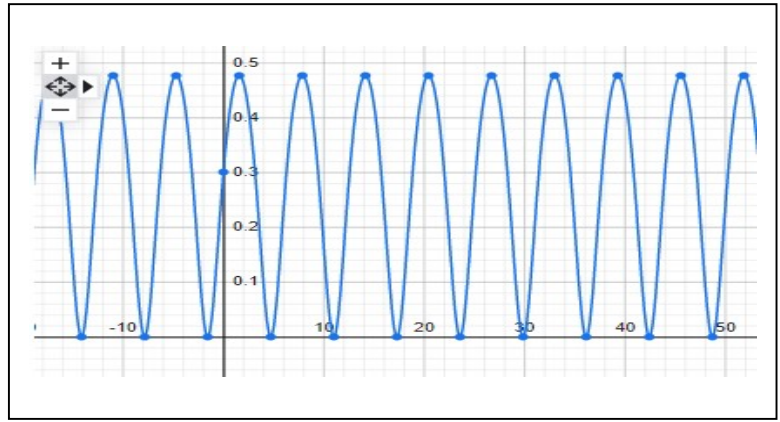

Figure 18 - Graph of the function $\log[\, cos\,(\omega t) + 2]$

Further, oscillations $y_r = a_2 \, cos(\omega_r t)$ (due to the nonlinearity of the tympanic membrane) with the resonant frequency of the middle ear cavity mix with the oscillations represented by the expression (8.2), resulting in:

$$\log y \cdot y_r = (\log A + c) \cdot a_2 \cos(\omega_r t) + a_1 \, cos(\omega t) \cdot a_2 \cos(\omega_r t) \,,$$

and write the product of functions as the basis of the transformation:

$$\log y \cdot y_r = a_2 (\log A + c) \cdot \cos(\omega_r t) + a_1 a_2 [cos((\omega_r + \omega) t) + cos((\omega_r - \omega) t)] \tag{7.3}$$

Thus, the stapes vibrates in accordance with expression (7.3). However, due to the fact that the cochlea is, let's say, a set of low-pass filters, only part of the expression is analyzed, namely:

$$\log y \cdot y_r = a_2 (\log A + c) \cdot \cos(\omega_r t) + a_1 a_2 [cos((\omega_r - \omega) t)] \tag{7.4}$$

So, from the perspective of radio engineering, we have a direct conversion superheterodyne receiver. Analyzing the obtained expression, we can say that the amplitude of the sound signal and its frequency are separated from each other. They can be calculated because their cosines have different frequencies, which is consistent with Heisenberg's uncertainty principle. In the cochlea, this will look like what is shown in Figure 11, where the first term of this expression will be reflected at the location with diameter d, and the other term of this sum - at the location with $d_i$ (attached to a location).

If the sound signal is a musical sound, and the second term is the sum of harmonics, i.e., the so-called harmonic wave, then this term can be written in the Fourier series as follows:

$$y_2 = a_1 a_2 [\sum_i \cos(\omega_r - \omega_i) t + \sum_i \sin(\omega_r - \omega_i) t], \quad \text{где i=1, 2, 3…} \tag{8.5}$$

However, this does not change the essence of the analysis because all the even-numbered frequencies will be located in one section of the corresponding diameter (as shown in Figure 19), and frequencies with an index i=2, 3... will be located in a section with a diameter corresponding to d/2. Frequencies with an index i=4, 5... will be located in a section with a diameter corresponding to d/4, and so on. However, due to the structure of the two membranes, not all frequencies will be perceived by this section. From Figure 19, it can be seen that even frequencies will not be perceived because their nodes, where the amplitude values of the standing wave are almost zero, are located between the two membranes.

Therefore, in the d section, frequencies with an index i=1, 3, 5... will be analyzed, in the d/2 section, frequencies with an index i=2, 6, 10..., and in the d/4 section, frequencies with an index i=4, 12, 20..., and so on. So, we have the first, second, and fourth formants, while the third can be calculated as the difference between the first and second formants. Therefore, it can be seen from this that the analysis of ear canal by musical sounds does not have significant problems because each musical sound has three phases: attack, stable zone, and decay. In the stable zone, which occupies the most significant time duration and during this time, it does not change its spectral composition, meaning its frequency parameters do not change.

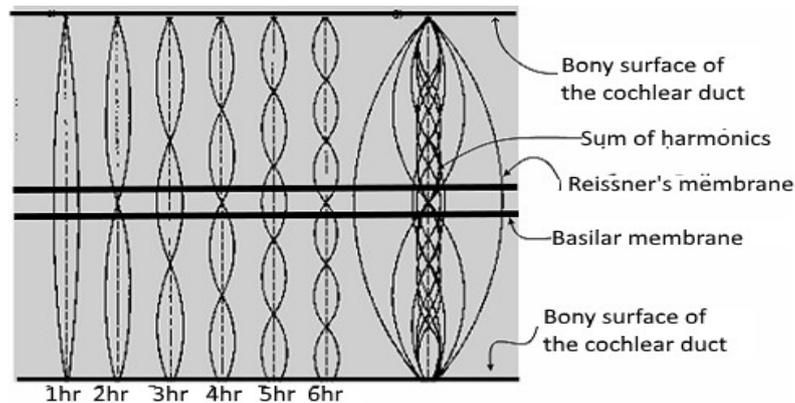

Figure 19 - Illustrations of the distribution of frequencies in the cross-section of the cochlea.

Note that musical sounds are produced by instruments with relatively stable but concentrated frequency characteristics in the stabilization zone, and furthermore, these sounds have only harmonic frequencies, such as a string. This means that their frequency spectrum is not blurred across the entire cochlear frequency range but is concentrated as much as possible along the lengths along the cochlea according to the law of the musical scale (see Figure 11). The actual series of cochlear frequencies has a coefficient much smaller than the musical scale itself, i.e. $\sqrt[m]{2} < \sqrt[12]{2}$, where $m$ is determined by the thickness of the transverse fibers of the basilar membrane, creating the discreteness of this series. The musical scale is an averaged band of frequencies on the basilar membrane, which varies for each individual. This allows us to analyze musical sounds even when their frequencies have some deviation or are not perfectly tuned. Considering the invariance of the musical scale in the frequency domain, we have only 12 notes because every musician knows that any melody can be played within one octave (the rest of the notes are harmonics of the fundamental octave). Furthermore, the fundamental pitch (the first harmonic or first formant) of the male voice ranges from 100-200 Hz, while for the female voice, it's within 150-300 Hz. In other words, the fundamental pitch of the voice changes by approximately one octave.

## 8. Vocal Tract

Language signals, or speech, are another matter. The sounds of speech are also a physical phenomenon. However, most researchers distinguish them from purely physical phenomena (such as the sounds of musical instruments, the clinking of metal, the noise of the wind, etc.) primarily because they are consciously produced by humans using their speech organs, making them a phenomenal physiological phenomenon. The point is that nature did not provide humans with a specific, separate organ for producing speech sounds. Instead, over the course of evolution, respiratory organs, digestive organs, the muscles of speech organs, and their entire functioning have adapted to this purpose, all under the control of the nervous system. However, physiology alone does not explain how we articulate our speech sounds so precisely.

It's worth noting that the speech signal has several characteristics that need to be considered: the signal's properties are not constant within the chosen word-length segment; it's a non-stationary random process; the signal's form is complex (speech is more like noise than a regular signal). To overcome these challenges, the discrete random process of digitized speech signals is considered quasi-stationary within an interval of approximately 10 ms because the parameters of the vocal tract do not change significantly over this interval. This time interval has been experimentally justified. Based on these premises, there are various methods for processing and modeling speech signals, including: spectral analysis on short time intervals and speech synthesis; homomorphic analysis and synthesis of speech signals; synthesis using linear prediction; modeling speech with transfer functions; waveform coding methods; parametric representation methods; synthesis according to specific rules [31]. The main drawback of these methods is that they consider the process of speech sound production as an open-loop system, i.e., without feedback. In reality, it's not that simple. We have a closed-loop monitoring system with feedback because our auditory system has two analyzers (right and left) in the form of cochlea and the corti organ, and no others have been found. These analyzers are used as measuring elements in the feedback loop. All the necessary conditions are in place for this. The middle ear cavity is connected to the pharyngeal (auditory) tube, which is an excellent waveguide. So, the sounds produced in the pharyngeal cavity, representing the vibrations created by the vocal cords, enter the middle ear cavity through the auditory tube, where they combine with the resonant frequency vibrations as follows:

$$y + y_r = \cos(\omega t) + \cos(\omega_r t) = 2\cos((\frac{\omega_r + \omega}{2})t)\,cos((\frac{\omega_r - \omega}{2})t),$$

and after logarithmization with the eardrum, we have:

$$\log(y + y_r) = \log 2 + \log(\cos((\frac{\omega_r + \omega}{2})t) + \log(\cos((\frac{\omega_r - \omega}{2})t).$$

Since our analyzer in the form of a cochlea acts as a low-pass filter, it only analyzes the difference in frequency:

$$\log(y + y_r) = \log 2 + \log(\cos((\frac{\omega_r - \omega}{2})t)),\qquad(8.1)$$

So, we hear ourselves an octave lower because of dividing this difference by two, that's why we sometimes don't recognize our own voice when we hear it in a recording.

Therefore, by using feedback when creating speech sounds, a person has the ability to adjust the frequency characteristics of these sounds. On the other hand, it is clear that the amplitude of the sound signal is not a parameter for encoding information, as the normalized signal is fed to the input of our analyzer through logarithmic transformation (Figure 18). So, the information is contained in the combination of frequencies. This is evident even from the fact that 12 musical sounds are clearly insufficient for encoding phonemes, of which there are more than 40 in Slavic languages. However, in the feedback loop, there is a significant delay (latency), which means that we cannot control each specific frequency separately because standing wave mode needs to be established for analysis in the cochlea. Corrections can be made in the subsequent phonemes, which the cochlea repeats more than once. This leads to non-stationarity of phonemes. But there is nothing wrong with this because our auditory analyzer recognizes not a single frequency but a certain frequency band allocated to a particular phoneme. Since a phoneme is repeated many times by a person, it doesn't matter how this band is filled over time for that sound. What matters is that it is filled. Thus, in this way, our analyzer deals with the non-stationarity of various phonemes.

## Conclusions

Most theories of hearing, including Rutherford's and Ewald's theories, deny the possibility of primary sound analysis in the cochlea. According to these claims, this is performed by higher auditory centers or systems, the structure of which is not entirely understood, and it is unknown where they are located, with no factual evidence to support this. The proposed technocratic model of hearing provides a clear positive answer to this question. We see that all primary sound analysis is entirely carried out by our ears, representing the results of the analysis in the form of distances that encode various spectral components. The neural network serving our auditory analyzer is exclusively involved in classification and recognition, which is its primary function. Therefore, it can be concluded that the cochlear labyrinth (the middle duct), consisting of a system of membranes and the Corti organ, is merely a data collection and amplification system along the cochlear spiral, serving as an interface between the analyzer and the neural network of the human brain.

Many authors also claim that the ear works as a bank of filters. However, they do not go beyond these claims because they cannot say where these filters are located and on what principles they work. The proposed theory of hearing can also be interpreted to some extent as a set of filters, but not in the traditional sense. Because each transverse section of the cochlea isolates a range of frequencies from the signal, which can be described by a discrete Fourier expression. Since the input sound signal in the cochlea is normalized through logarithmization, meaning the amplitude of each spectral component is fixed, such a series is called Fourier descriptors and is widely used for training in neural networks. Therefore, the expression can be rewritten as:

$$y_2 = a_1 a_2 [\sum_n \cos(n \omega t) + \sum_n \sin(n \omega t)] = \sum_n DF_n(n \omega), \qquad (9.1)$$

where $\omega = (\omega_r - \omega_i)$ is the fundamental frequency of the standing wave in the cochlea (first harmonic), n=1, 2, 3,..

Thus, expression (9.1) describes our filter, and then the speech signal can be represented as:

$$y(t) = \sum_{\omega_m = \omega_1}^{\omega_m = \omega_2} \sum_n DF_{n,\omega_m}(n \omega_m), \qquad (9.2)$$

де $\square_1, \square_2$ are the boundaries of the frequency band occupied by the speech signal, $\omega_n$ is the frequency within this band that changes according to the law $\omega_{n+1} = \sqrt[12]{2} \cdot \omega_n$.

The nonlinear nature of frequencies in the specified band, on one hand, is positive because the frequencies take on values of infinite fractions, which means they are non-integer multiples. This implies that they are independent and easily distinguishable without requiring the constraints imposed in other cases by Shannon's theorem. On the other hand, for this reason, classical Fourier spectral analysis does not yield the desired results in speech recognition.

It should also be noted that the phenomenon of transverse resonance in most technical systems is a parasitic phenomenon that interferes with the transmission of information in waveguides. However, evolution has made it positive and useful in the context of the cochlea, but it requires more detailed acoustic research into the properties of tapered, coiled tubes.

Further research is also needed on the pulsatile wave in the Reissner's membrane, as many hearing impairments depend to a large extent on its quality and elasticity.

It may also give the impression that the proposed model of human hearing contradicts Bekesy's theory, rejecting it. However, in our theory, the traveling wave of Bekesy in the basilar membrane is merely an illusion created by standing transverse waves following the known principle of "traveling fires" with the appropriate phase shift. Therefore, we have a non-standard way of wave analysis in which any traveling wave can be decomposed into a series of standing transverse waves with the corresponding phase shift, and vice versa, a traveling wave can be created by standing waves with the same named shift.

# References

1. Campbell N. A., Reece J. B. *Biology*. 8th ed. Benjamin Cammings. 2008. Archived from the original on March 3, 2011. Cited on June 19, 2011.

2. Marieb E. N., Hoehn K. Human Anatomy & Physiology. 7th ed. 2006.

3. Guest J. F., Greener M. J., Robinson A. C., Smith A. F. Impacted cerumen: composition, production, epidemiology and management. *QJM*. 2004 Aug. 97 (8). P. 477-88. doi:10.1093/qjmed/hch082.

4. Netter, F. Atlas of Human Anatomy. Edited by Prof. Yu. B. Chaykovsky; translated from English by A. A. Tsegelsky, Ph.D. Lviv: Nautilus, 2004. 592 pages.

5. Taylor D., Green N., Stout W. Biology: in 3 volumes. Translated from English under the editorship of R. Soper. 3rd edition. Moscow: Mir, 2004. Volume 2. 436 pages.

6. Hickman C. P. Jr., Roberts L. S., Larson A. Integrated principles of zoology. 11th ed. McGraw-Hill Higher Education, 2001.

7. Prosser C. L., Bishop D. V., Brown F. A. Jr., Jahn T. L., Wulf V. J. Comparative animal physiology. W.B. Saunders Company, 1950.

8. Encyclopædia Britannica. Archived from the original on July 3, 2011. Cited June 18, 2011.

9. Biology: a textbook / edited and translated from Russian by V.O. Motuznyi. 3rd edition, revised and expanded. K.: Vysha Shkola, 2002. 622 pages.

10. Helmholz H.L.F. Die Lehre von Tonempfindungen. Draunschweig, 1863 / пер. Петрова. СПб, 1875.

11. Guide to Otorhinolaryngology / ed. by I. B. Soldatov. B. Soldatov. Moscow: Medicine, 1994.

12. Hearing: An Introduction to Psychological and Physiological Acoustics/ translated from English and published by "Medicine" in Moscow,1984.

13. Bekesy G. Experiments in Hearing. New York: McGraw-Hill, 1960.

14. Shuplyakov V. S. Collection: Sensory Systems. Hearing. Leningrad: Science, 1982. Pp. 3-17.

15. Human Physiology, edited by G. I. Kositsky. 3rd edition. Moscow: Medicine, 1985.

16. Основы физиологии / под ред. П. Стерки. М.: Мир, 1984.

17. Физиология человека : Compendium / под ред. Б. И. Ткаченко и В. Ф. Пятина. Спб, 1996.

18. Human Anatomy and Physiology / 3rd ed. E. N. Marieb. The Benjamin Commings Publishing Company, Inc. California, 1995.

19. The World's Best Anatomical Charts / Anatomical Chart Co. Skokle, IL, 1993.

20. Tamar G. Fundamentals of Sensory Physiology, translated from English. Moscow: Mir, 1976.

21. Prives M. G., Lysenko N. K., Bushkovich V. I. Human Anatomy, edited by M. G. Prives. Moscow: Medicine, 1985.

22. Histology, edited by V. G. Eliseev, Yu. I. Afanasyev, E. A. Yurina. Moscow: Medicine, 1983.

23. Rutherford W. *J. Anat. Physiol*. 1986. Vol. 21. P. 166-168.

24. Wever E. G. Theory of Hearing. New York: Dover, 1949.

25. Ovchinnikov E. L., Eremina N. V. Method for detecting biophysical processes implementing the mechanism and biophysical (wave) model of human hearing. Russian Federation Patent.

26. Reichenbach T., Stefanovic A., Nin F., Hudspeth A.J. Waves on Reissner's Membrane: A Mechanism for the Propagation of Otoacoustic Emissions from the Cochlea / Howard Hughes Medical Institute and Laboratory of Sensory Neuroscience, The Rockefeller University, New York, NY 10065-6399, USA. doi: 10.1016/j.celrep.2012.02.013

27. Acoustics of Hearing: Lecture Notes. Textbook for Students of Specialty 171 'Electronics,' Educational Program 'Acoustic Electronic Systems and Acoustic Information Processing Technologies.' Compiled by K. S. Drozdenko, O. I. Drozdenko. Electronic Text Data. Kyiv: Igor Sikorsky Kyiv Polytechnic Institute, 2020. 99 pages. URL: https://ela.kpi.ua/bitstream/123456789/35008/1/Akustyka.pdf (Last accessed: 11.09.2023).

28. Multimode Waveguides with Random Irregularities" by R.B. Vaganov, R.F. Matveev, V.V. Meriakri. Published by Sov. Radio in 1972. 232 pages.

29. Hearing and Sound Perception: Lecture Course" by E.I. Vologdin. Published by "ST Faculty DVO" in St. Petersburg in 2004. 36 pages.

30. L. M. Fink. Signal Transmission Theory. Moscow: Svjaz, 1980. 288 pages.

31. O. N. Karpov. Methods of Analysis and Recognition of Speech Signals. Dnipro, 2001. Pp. 25-36.